\documentclass[10pt,letterpaper]{article}
\usepackage[top=0.85in,left=1.75in,footskip=0.75in,marginparwidth=2in]{geometry}

\usepackage[utf8]{inputenc}

\usepackage{cite}

\usepackage{nameref}
\usepackage{hyperref}
\usepackage{xcolor}
\hypersetup{
    colorlinks,
    linkcolor={red!60!black},
    citecolor={green!60!black},
    urlcolor={blue!80!black}
}

\usepackage[right]{lineno}

\usepackage{microtype}
\DisableLigatures[f]{encoding = *, family = * }

\raggedright
\usepackage{changepage}

\usepackage[aboveskip=1pt,labelfont=bf,labelsep=period,singlelinecheck=off]{caption}

\makeatletter
\renewcommand{\@biblabel}[1]{\quad#1.}
\makeatother

\usepackage{lastpage,fancyhdr,graphicx}
\usepackage{epstopdf}
\fancyheadoffset[L]{2.25in}
\fancyfootoffset[L]{2.25in}

\usepackage{color}

\definecolor{Gray}{gray}{.25}

\usepackage{graphicx}

\usepackage{sidecap}

\usepackage{wrapfig}
\usepackage[pscoord]{eso-pic}
\usepackage[fulladjust]{marginnote}
\reversemarginpar

\begin{document}
\vspace*{0.35in}

\begin{flushleft}
{\Large
\textbf\newline{Determining JPEG Image Standard Quality Factor from the Quantization Tables.}
}
\newline
\\
Rémi Cogranne
\\
\bigskip
Troyes University of Technology, Lab. for System Modelling and Dependability,\\
ROSAS Dept., ICD, UMR 6281 CNRS, Troyes 10010 Cedex, CS 42060, France
\\
\bigskip
* remi.cogranne@utt.fr

\end{flushleft}

\section*{Abstract}
Identifying the quality factor of JPEG images is very useful for applications in digital image forensics. 
Though several command-line tools exist and are used in widely used software such as \emph{GIMP} (GNU Image Manipulation Program), the well-known image editing software, or the \emph{ImageMagick} suite, we have found that those may provide inaccurate or even wrong results. 
This paper presents a simple method for determining the exact quality factor of a JPEG image from its quantization tables. 
The method is presented briefly and a sample program, written in Unix/Linux Shell bash language is provided.

\section{Introduction}
The JPEG compression standard, named after the Joint Photographic Experts Group that proposed this scheme, has been very widely adopted since it publication in 1993. It is indeed a computation simple, a highly flexible and a rather efficient compression scheme that allows easily setting the compression rate, raging from artifacts invisible with naked eyes to harsh compression that can slash the image size by 50 or even more. 

The two main parts of JPEG lossy compression scheme are the color subsampling and Discrete Cosine Transform (DCT). An uncompressed image is first subjected to color transcoding, from RGB images to YCbCr luminance/chrominance colorspace, followed by chroma subsampling which remove half of the total number of samples using the usual 4:2:0 subsampling system. Each channel is then processed block-wise, with blocks of size of $8\times8$ pixels ; over each block the DCT is applied. The resulting coefficients are quantized with different quantizations steps to keep more information on low frequencies components. The quantization steps of all coefficients, some referred to as modes or frequencies, are defined in a so-called quantization matrix, or sometimes referred to as quantization tables (of course of size $8\times8$).\\
Those two main parts of the lossy JPEG compression scheme are based on studies that show how the Human Visual System (HVS) is more sensitive to luminance than chroma, hence keeping more information on luminance, and more sensitive to lowest frequencies that represents the rough content rather than highest frequencies which represents small details. For a complete description on the JPEG compression scheme, the reader may refer to the article~\cite{wallace1992jpeg} that provides a detailed introduction or the complete reference book~\cite{pennebaker1992jpeg} that provides all the details and even reproduces the whole standard~\cite{JPEGnorm}.

The quantization matrix of a JPEG images is provided in the header of the file so that the decoder knows each quantization step for each DCT coefficient. There are two types of quantization matrix: (a) the JPEG norm~\cite[Annex K.1]{JPEGnorm} provides a standard way to obtain a quantization matrix given a quality factor that is chosen between 1 (lowest quality, all quantization steps equal 255) and 100 (highest quality, all quantization steps equal 1), and, (b) non-standard quantization matrix, the JPEG norm accepts any quantization step between $1$ and $255$ as elements of the quantization matrix. One can therefore design its own quantization matrix, this is the case is many digital cameras and several image editing software, such as Adobe\textregistered Photoshop and Lightroom.

Over the past decades, many image forensics works exploited the quantization tables to detect forgeries or identify image origin~\cite{farid2008digital,Kornblum2008,Babak2011QTforensics,Stamm2013forensics_review,Thai2015}. 
Some images forensics tools also exploits the JPEG quantization matrix such as the 
\emph{JPEGsnoop} project\footnote{\emph{JPEGsnoop} source code is available on GitHub at \href{https://github.com/ImpulseAdventure/JPEGsnoop}{github.com} along with its documentation from \href{http://www.impulseadventure.com/photo/jpeg-snoop.html}{www.impulseadventure.com}.}, which uncover a large set of information from the JPEG image file header that can be cross-checked from the project database.\\
Similarly many prior works on image forensics focused on estimating the quantization matrix 
from either JPEG images compressed twice or stored back in uncompressed format~\cite{fridrich2008detection,luo2010jpeg,THT2017Quantization_estimation} and those works usually assumed that the quantization matrix is one of the standard ones. Quantization matrices are also very useful in information hiding. The use of quantization matrix has been exploited in steganography with side-information~\cite{Denemark2017MBS} and most recent steganalysis~\cite{Thai2013ICIP,Holub2015DCTR,Song2015GFR,Denemark2016SCA} are tailored for standard quantization matrices. The recent work~\cite{Giboulot2018Wild} show how important the quantization tables are for information hiding detection. Similarly, statistical models of DCT coefficients~\cite{Thai2012ICIP,Thai2013TIP} take into account the quantization tables and have been assessed with standard quantization matrices.

\subsection{Contribution of the Proposed Method}\label{ssec:intro_contrib}

However we noticed that is currently no tools for checking whether a JPEG images has been compressed using a standard quantization matrix. Even worst, several tools may provide inaccurate or even wrong results. For instance, \emph{JPEGsnoop} returns an ``\textrm{Approx quality factor = 74.75}'' for a JPEG compressed with standard quantization matrix from quality factor 75 and an ``\textrm{Approx quality factor = 97.68}'' when using the standard quantization matrix from quality factor 95. Similarly, digging into the source code of the well-known  \emph{ImageMagick} software suite determine the compression factor, from the quantization tables, using only the sum of all the quantization steps\footnote{See the file \href{https://github.com/ImageMagick/ImageMagick/blob/05d2ff7ebf21f659f5b11e45afb294e152f4330c/coders/jpeg.c\#L818}{jpeg.c from ImageMagick Source code}.}. Such estimation is obviously inaccurate and, though unlikely, can return find an exact quality factor while the quantization matrices do not match the standard one if the sum of all elements does.\\
This is partially explained by the fact that the comparison of all the 128 quantization steps (for luminance and chrominance quantization matrices) with all the possible quantization matrices is time consuming.

This paper describes a simple, efficient and very accurate method for checking whether the quantization matrices of a given JPEG images match the standard ones and, if so, to identify the associated quality factor. The proposed method is based on a step approach. First the possible quality factors are identified from the quantization tables. This first step often returns a single possible quality factor, and in several cases two, from which the exact match can be checked quickly simply by comparing all the coefficients one by one.\\
Though this approach is not very innovative on a scientific point of view, we publish a paper describing the method associated with a sample Shell source code, using the Bash programing language, since it can be useful for the research community.

\begin{table}[t]
\begin{center}
\begin{tabular}{|c|c|c|c|c|c|c|c|}\hline
16	&	11	&	10	&	16	&	24	&	40	&	51	&	61	\\\hline
12	&	12	&	14	&	19	&	26	&	58	&	60	&	55	\\\hline
14	&	13	&	16	&	24	&	40	&	57	&	69	&	56	\\\hline
14	&	17	&	22	&	29	&	51	&	87	&	80	&	62	\\\hline
18	&	22	&	37	&	56	&	68	&	109	&	103	&	77	\\\hline
24	&	35	&	55	&	64	&	81	&	104	&	113	&	92	\\\hline
49	&	64	&	78	&	87	&	103	&	121	&	120	&	101	\\\hline
72	&	92	&	95	&	98	&	112	&	100	&	103	&	99	\\\hline
\end{tabular}
\hspace*{2cm}
\begin{tabular}{|c|c|c|c|c|c|c|c|}\hline
17	&	18	&	24	&	47	&	99	&	99	&	99	&	99	\\\hline
18	&	21	&	26	&	66	&	99	&	99	&	99	&	99	\\\hline
24	&	26	&	56	&	99	&	99	&	99	&	99	&	99	\\\hline
47	&	66	&	99	&	99	&	99	&	99	&	99	&	99	\\\hline
99	&	99	&	99	&	99	&	99	&	99	&	99	&	99	\\\hline
99	&	99	&	99	&	99	&	99	&	99	&	99	&	99	\\\hline
99	&	99	&	99	&	99	&	99	&	99	&	99	&	99	\\\hline
99	&	99	&	99	&	99	&	99	&	99	&	99	&	99	\\\hline
\end{tabular}
\end{center}
\caption{The two standard quantization matrices provided by the Independent JPEG Group (IJG).}
\label{tbl:def_quantization_matrices}
\end{table}

\begin{table}[t]
\begin{center}
\begin{tabular}{|c|c|c|c|c|c|c|c|}\hline
 8 &  6 &   5 &   8 &   12 &   20 &  26 &   31 	\\ \hline
 6 &   6 &   7 &   10 &   13 &   29 &   30 &   28 	\\ \hline
 7 &   7 &   8 &   12 &   20 &   29 &   35 &   28 	\\ \hline
 7 &   9 &   11 &   15 &   26 &   44 &   40 &   31 	\\ \hline
 9 &   11 &   19 &   28 &   34 &   55 &   52 &   39 	\\ \hline
 12 &   18 &   28 &   32 &   41 &   52 &   57 &   46 	\\ \hline
 25 &   32 &   39 &   44 &   52 &   61 &   60 &   51 	\\ \hline
 36 &   46 &   48 &   49 &   56 &   50 &   52 &   50 	\\ \hline
\end{tabular}
\hspace*{2cm}
\begin{tabular}{|c|c|c|c|c|c|c|c|}\hline
 9 &   9 &   12 &   24 &   50 &   50 &   50 &   50 	\\ \hline
 9 &   11 &   13 &   33 &   50 &   50 &   50 &   50 	\\ \hline
 12 &   13 &   28 &   50 &   50 &   50 &   50 &   50 	\\ \hline
 24 &   33 &   50 &   50 &   50 &   50 &   50 &   50 	\\ \hline
 50 &   50 &   50 &   50 &   50 &   50 &   50 &   50 	\\ \hline
 50 &   50 &   50 &   50 &   50 &   50 &   50 &   50 	\\ \hline
 50 &   50 &   50 &   50 &   50 &   50 &   50 &   50 	\\ \hline
 50 &   50 &   50 &   50 &   50 &   50 &   50 &   50 	\\ \hline
\end{tabular}
\end{center}
\caption{Ensuing quantization matrices for quality factor $98$.}
\label{tbl:QM98}
\end{table}

\begin{table}[t]
\begin{center}
\begin{tabular}{|c|c|c|c|c|c|c|c|}\hline
 1 &    1 &   1 &   1 &   1 &   2 &   2 &   2 	\\ \hline
 1 &   1 &   1 &   1 &   1 &   2 &   2 &   2 	\\ \hline
 1 &   1 &   1 &   1 &   2 &   2 &   3 &   2 	\\ \hline
 1 &   1 &   1 &   1 &   2 &   3 &   3 &   2 	\\ \hline
 1 &   1 &   1 &   2 &   3 &   4 &   4 &   3 	\\ \hline
 1 &   1 &   2 &   3 &   3 &   4 &   5 &   4 	\\ \hline
 2 &   3 &   3 &   3 &   4 &   5 &   5 &   4 	\\ \hline
 3 &   4 &   4 &   4 &   4 &   4 &   4 &   4 	\\ \hline
\end{tabular}
\hspace*{2cm}
\begin{tabular}{|c|c|c|c|c|c|c|c|}\hline
 1 &   1 &   1 &   2 &   4 &   4 &   4 &   4 	\\ \hline
 1 &   1 &   1 &   3 &   4 &   4 &   4 &   4 	\\ \hline
 1 &   1 &   2 &   4 &   4 &   4 &   4 &   4 	\\ \hline
 2 &   3 &   4 &   4 &   4 &   4 &   4 &   4 	\\ \hline
 4 &   4 &   4 &   4 &   4 &   4 &   4 &   4 	\\ \hline
 4 &   4 &   4 &   4 &   4 &   4 &   4 &   4 	\\ \hline
 4 &   4 &   4 &   4 &   4 &   4 &   4 &   4 	\\ \hline
 4 &   4 &   4 &   4 &   4 &   4 &   4 &   4 	\\ \hline
\end{tabular}
\end{center}
\caption{Ensuing quantization matrices for quality factor $75$.}
\label{tbl:QM75}
\end{table}

\section{Identification of Quality Factor}

\subsection{Determing Possible Quality Factors}

Before describing the proposed method for identifying the possible quality factor, let us recall the relation between the quality factor $F$ and the ensuing quantization matrices $\mathbf{Q}$ ; note that for clarity the method is described only for the quantization of the luminance channel since it works exactly the same way for chrominance.\\ 
For a given quality factor $F$, the elements of associated quantization matrix, denoted $\forall (i,j) \in \{1,8\}^2 \,,\ Q_{i,j}$, also referred to as the quantization steps, are obtained using the following relation:
\begin{equation}
	Q_{i,j} = \left[\frac{50 + S \times D_{i,j} }{100}\right] ,
\label{eq:quantization_mat_def}
\end{equation}
where $[x]$ denotes the rounding (quantization) of variable $x$, $D_{i,j}$ are the elements of default quantization matrices provided by the IJG (Independent JPEG Group) and where the parameter $S$ is given by $S=200-2F$ in the very usual case of quality factor $F>=50$. Otherwise, the parameter $S$ is given by $S=5000/Q$ ; this allows an increase of quantization steps $Q_{i,j}$, see~(\ref{eq:quantization_mat_def}), at a much faster pace when $F$  decreases.

The default quatization matrices from IJG are reported in Table~\ref{tbl:def_quantization_matrices} and, as examples, the ensuing quantization matrices for quality factors $F=98$ and $F=75$ are provided in Tables~\ref{tbl:QM98} and~\ref{tbl:QM75} respectively.\\
We used the Equation~(\ref{eq:quantization_mat_def}) to identify possible quality factors, from quantization matrices, by determining the range of possible values of parameters $S$ for all quantization steps. From relation~(\ref{eq:quantization_mat_def}) it immediately follows that:
\begin{equation}
	Q_{i,j}-\frac{1}{2} \leq \frac{50 + S \times D_{i,j} }{100}  < Q_{i,j}+\frac{1/2}.
\label{eq:identify_QF1}
\end{equation}
for the usual definition of quantization, that is $[x] = \lfloor x + 1/2 \rfloor$ with $\lfloor \cdot \rfloor$ the flooring function (integer part).\\
However, some software may use different quantization function, therefore we enlarge to the interval to $[y]x \Rightarrow    x-1 \leq y \leq x+1$ which, applied to relation~(\ref{eq:quantization_mat_def}), leads to :
\begin{equation}
	Q_{i,j}-1 \leq \frac{50 + S \times D_{i,j} }{100}  < Q_{i,j}+1.
\label{eq:identify_QF_large}
\end{equation}



It immediately follows from Equation~(\ref{eq:identify_QF_large}) that, for each quantization step $Q_{i,j} \,,\ 1\leq i \leq 8 \,,\ 1\leq j \leq 8$, the parameter $S$ can be bounded as follows:
\begin{equation}
	\frac{ 100 \times Q_{i,j} - 150}{D_{i,j}} \leq S  <  \frac{ 100 \times Q_{i,j} + 50}{D_{i,j}}.
\label{eq:identify_QF_final}
\end{equation}

The principle of the proposed approach is to use the bounds given in Equation~(\ref{eq:identify_QF_final}) over each quantization step, $Q_{i,j} \,,\ 1\leq i \leq 8 \,,\ 1\leq j \leq 8$,to narrow te bounds as much as possible.\\
One eventually gets tight bounds from which the possible quality factor(s) can be obtained using the relation
\begin{equation}
	F=\frac{200-S}{2}
\label{eq:from_S_to_F}
\end{equation}
Note that this relation is only valid for quantity factor greater than 49. We only considered this case, since JPEG images with quality factor smaller are extremely rare. However a simple adaptation for $S$ greater than $100$ is straightforward.

As already stated at the end of Section~\ref{ssec:intro_contrib}, this first step very often returns a single possible quality factor. Therefore, using this result, one can quickly check, simply by comparing all the coefficients one by one, the exact match between the possible quality factor and the associated standard quantization matrix.\\

\subsection{Practical Implementation}

The source code attached along with this paper provides a simple implementation of the approach described using the Shell  Bash programming language. 
The only external tool used in that sample code is \emph{exiftool} by Phil Harvey~\footnote{The command-line tool \emph{exiftool} is available from author its \href{http://www.sno.phy.queensu.ca/~phil/exiftool/}{author's, Phil Harvey, webpage}.} which allows extracting the quantization matrices from a JPEG image (using the option \textrm{-v3}). Alternative, one can simply parse the JPEG image file and seeks the \textrm{DQT} (Define Quantization Tables) marker that can be easily identified as with its corresponding code \texttt{0xFFDB} in hexadecimal notation~\cite[Annex B]{JPEGnorm}.\\

The code allows to check the quality factor using either only the luminance channel quantization matrix, only the chrominance quantization matrix or both. This option is specified by the first argument. This argument can therefore be either $1$, $2$, or $3$, coding in bits ($01$, $10$ or $11$) the use of luminance and chrominance channel respectively.

A second argument control the display of the script. This second argument can be either $0$ for no output, in such case the return code (or exit status) to report the quality factor (between $1$ and $100$) ; in case of error, return codes greater than $100$ are output ($200$ for JPEG image file not found, $101$ for no possible matching quality factor and $102$ when a candidate quality factor is found, but the check of all quantization steps do not match).\\
When the second argument is set to $1$ a single line display the results of the determined quality factor with the return code, or exit status, being also used. \\
Eventually a second argument of $2$ makes the script verbose, displays the quantization tables, the candidate quality factor found and the result of quantization steps check. 

Examples of the use of the script are given below :
\begin{description}
  \item[\$] \texttt{./JPEG\_QF\_identify2.sh ./image.jpg 3 1} : determine the quality factor using both luminance and chrominance channels quantization matrices with single line output.
  \item[\$] \texttt{./JPEG\_QF\_identify2.sh ./image.jpg 1 2} : determine the quality factor using only chrominance channel quantization matrix with verbose output.
  \item[\$] \texttt{./JPEG\_QF\_identify2.sh ./image.jpg 1 0} : determine the quality factor using only luminance channel quantization matrix without output (results in return code, or exit status, that is in the variable \texttt{\$?}.
\end{description}

\section{Results}

First and most important, we have tried the script provided along with this paper on images with all quality factors between $50$ and $100$, included, saved using the \emph{convert} command-line tool, that is a part of \emph{ImageMagick} suite, \emph{Matlab} numerical computing software with the command \emph{imwrite} and the \emph{GIMP} image editing software. With all those images, using the chrominance channel, the luminance channel or both, we have noted any error on quality factor identification. Besides, we have randomly selected images and modified each time a single quantization step (this has been done using the hexadecimal text editor \emph{jeex}); as expected in all those cases, the proposed script has been able to identify a non-standard quantization matrix, sometimes by not finding any candidate quality factor, but in most of the cases by detecting different quantization steps with the candidate quality factor. 

Though the Bash shell programming language is not very efficient, the script has been found rather efficient. Using the Linux \emph{time} command, we measured an average running time of about $0.497$s when using both chrominance and luminance channels. This can be compared with the average running time of about $0.318$s when using only the luminance channel and $0.317$s when using only the chrominance channel. Note that those running times are averaged over $255$ images (that is, repeated five times over all images with standard quantization matrices and quality factors between $50$ and $100$). This corresponds to an increase of about $56.7$\% of the running time when using both luminance and chrominance quantization matrices. Interestingly, we note that the ``user'' running time is increased by about $50.5$\% as opposed to a ``system'' running time (time spent in the system kernel) of about $98$\%.

\section{Statement}

The provided source code is provided ``as it'' without any guarantee nor liability under license CC-BY-NC\footnote{see more detail on creative common webpage on  \href{https://wiki.creativecommons.org/wiki/License_Versions}{License Versions} and especially on \href{https://creativecommons.org/licenses/by-nc/4.0/legalcode}{Licence CC-BB-NC 4.0}.} which, in brief, allows to use or reuse this source code except for commercial purposes, provided that credits to the present works are given.

\nolinenumbers

%

\end{document}